\documentclass[journal=jacsat,manuscript=article]{achemso}

\usepackage[version=3]{mhchem} % Formula subscripts using \ce{}
\usepackage{color,soul}

\author{Han Zhang}
\affiliation[University of Pennsylvania]
{Department of Chemical and Biomolecular Engineering, University of Pennsylvania, Philadelphia, PA 19104, USA}
\author{Robert A. Riggleman}
\email{rrig@seas.upenn.edu}
\affiliation[University of Pennsylvania]
{Department of Chemical and Biomolecular Engineering, University of Pennsylvania, Philadelphia, PA 19104, USA}

\title[An \textsf{achemso} demo]
  {Predicting failure locations in model end-linked polymer networks}

\abbreviations{IR,NMR,UV}

\begin{document}

\begin{tocentry}

Some journals require a graphical entry for the Table of Contents.
This should be laid out ``print ready'' so that the sizing of the
text is correct.

Inside the \texttt{tocentry} environment, the font used is Helvetica
8\,pt, as required by \emph{Journal of the American Chemical
Society}.

The surrounding frame is 9\,cm by 3.5\,cm, which is the maximum
permitted for  \emph{Journal of the American Chemical Society}
graphical table of content entries. The box will not resize if the
content is too big: instead it will overflow the edge of the box.

This box and the associated title will always be printed on a
separate page at the end of the document.

\end{tocentry}

\begin{abstract}
The fracture of end-linked polymer networks and gels has a significant impact on the performance of these versatile and widely used materials, and a molecular-level understanding of the fracture process is crucial for the design of new materials. Network analysis techniques, especially geodesic edge betweenness centrality (GEBC) have been proven effective in failure locations across various network materials. In this work, we employ a combination of coarse-grained molecular dynamics simulations and network analysis techniques to investigate the effectiveness of GEBC and polymer strand orientation in predicting failure locations in model end-linked polymer networks. We demonstrate that polymer strands with fewer topological defects in their local surroundings, higher GEBC values compared to the system average, and greater alignment to the deformation axis are more prone to breaking under the uniaxial tensile deformation. Our results can be used to further refine the description of the processes at play during the failure of polymer networks and provide valuable insights into the inverse design of network materials with desired fracture properties.
\end{abstract}

\section{Introduction}
% Importance of polymer networks and Fracture
Polymer networks and gels are significant and highly versatile materials with their broad applications across diverse fields such as membranes, drug delivery systems and soft electronic devices. \cite{rogersMaterials2010, ionovHydrogelbased2014, liDesigning2016, guPolymer2020, danielsenMolecular2021, zhaoSoft2021, zhou3D2023} Given their wide-ranging use in people's daily lives, the fracture of these soft materials profoundly impacts their applications and performance. Understanding the fracture of polymer networks from a molecular perspective is crucial for advancing their utilization and facilitating the design of new materials. \cite{tauberStretchy2022}

% Role of topology and defects
Investigating the relationship between the macroscopic properties of polymer networks and their network structures and topology has been a challenging and long-standing problem in the fields of polymer physics and soft matter. \cite{creton50th2017} The properties of these soft amorphous materials depend on how molecules connect with each other in the networks, and the existence of topological defects, including loops and dangling ends, further complicates the understanding of these relationships. Over the last decade, the impact of topological defects on the polymer network elasticity \cite{zhouCounting2012, zhongQuantifying2016, wangUniversal2016, langElasticity2018, panyukovLoops2019, linRevisiting2019} and fracture \cite{linFracture2020, aroraFracture2020, aroraCoarseGrained2022, barneyFracture2022} has been incorporated into classic models, \cite{rubinsteinPolymer2003, lakeStrength1967} 
resulting in improved agreement with experimental results. Despite these advancements, the relationship between the fracture behavior and the molecular-level structures of polymer networks still remains elusive.

% Network analysis
Network analysis has recently emerged as a novel approach for studying complex disordered systems.  \cite{newmanNetworks2018, bassettInfluence2012, papadopoulosNetwork2018, berthierForecasting2019, kollmerBetweenness2019, pournajarEdge2022, nabizadehStructure2022a, mangalTopological2023, fazelpourCommunity2023} Representing disordered particulate systems as network structures offers an approach to characterize those systems across multiple length scales. Among various metrics used to quantify the network structures, the geodesic edge betweenness centrality (GEBC) has proven to be particularly useful in predicting failure locations in various disordered systems. \cite{berthierForecasting2019, morettiNetwork2019a, pournajarEdge2022, mangalTopological2023} GEBC, a specific type of betweenness centrality, measures the degree to which an edge lies on the shortest (geodesic) paths connecting nodes within the networks \cite{freemanSet1977, newmanNetworks2018}.  Berthier et al. demonstrated the effectiveness of GEBC in accessing possible failure locations in 2-dimensional disordered lattices with structures constructed from contact networks in granular media \cite{berthierForecasting2019}. Mangal et al. found that bonds with higher GEBC are more likely to rupture in short-ranged weakly attractive colloidal gels at different deformation rates \cite{mangalTopological2023}. These studies highlight the utility of the network analysis techniques, especially GEBC as a predictive tool for understanding the failure behavior of diverse disordered systems. While molecular simulations have become increasingly common tools to study polymer networks in recent years, \cite{grestKinetics1992, dueringStructure1994, wangUniversal2016, nowakTuning2016, nowakOptimizing2017, gusevNumerical2019, langAnalysis2020a, yeMolecular2020, barneyFracture2022, aroraCoarseGrained2022, zhangPercolation2023, masubuchiPhantom2023, sorichettiStructure2023} there still exists a gap in our understanding between network topology and properties. Network analysis techniques offers a promising route to incorporate chain-level information to the existing polymer network models and therefore holds immense potential for unraveling the intricate interplay between their molecular structures and macroscopic behaviors.
% This paragraph needs some kind of closing statement of the outstanding/unsolved problem

%In the context of network analysis, GEBC offers a way to identify "bridges" connecting different components in a graph, making it potentially valuable for identifying those "bridging strands" during the crack propagation in polymer networks.

% This work
In this work, we combine molecular dynamics simulations and network analysis techniques to predict the failure locations in model end-linked polymer networks with topological defects. By analyzing the isoconfigurational ensemble with uniaxial tensile deformations, we find that the failure locations of polymer networks is influenced by the underlying network structure. We find that for polymer strands with high probabilities of breaking, their local environment contains fewer primary loop defects compared to those with low probability of breaking. The presence of the topological defects also leads to a non-uniform distribution of GEBC across polymer strands within the networks, enabling the prediction of failure locations. GEBC of each strand and the angle between the deformation axis and each strand are calculated from the initial undeformed states of the networks. We find a positive correlation between GEBC and the probability of breaking of polymer strands and the strands that are more aligned with the deformation axis have higher probability of breaking. Moreover, strand breaking events initiate from polymer strands with high GEBC values and small angles between the strands and the deformation axis. These findings provide an effective approach to identify potential failure locations within the polymer networks based solely on their initial undeformed configurations, offer a novel perspective to understand the impact of topological defects in the networks and shed lights on the inverse design strategies of network materials with desired fracture properties.

\section{Methods}
\subsection{Network formation, equilibration and deformation}
We implement coarse-grained molecular dynamics (MD) simulations in the LAMMPS package \cite{plimptonFast1995, thompsonLAMMPS2022} to generate end-linked polymer networks following our previous works. \cite{yeMolecular2020, barneyFracture2022} Our approach aims at mimicking experimental systems where linear polymer chains with reactive endgroups react with tetrafunctional crosslinkers \cite{walkerWide2014}, and previous comparisons of our simulations to experimental results shows very good qualitative agreement in their properties. \cite{barneyFracture2022}

% Soft push-off and equilibration
We begin with a polymer melt of $2n$ identical mono-dispersed polymer chains with $N$ monomers per chain and $n$ crosslinkers that are chemically identical to the polymer monomers. In this work, we study systems with $N = 5, 15$ and $50$ and we keep the number of monomers $n_{monomer} = 2nN = 80,000$ the same across different systems ($n_{monomer} = 80,010$ for the $N = 15$ system to keep an integer number of chains). Polymer chains and crosslinkers are placed randomly into the simulation box, with periodic boundary conditions applied in all dimensions. The soft push-off method is employed first to remove any particle overlaps. \cite{auhlEquilibration2003, sliozbergFast2012} All crosslinkers and monomers are then modeled as chemically identical Lennard-Jones (LJ) beads.  Non-bonded interactions are modeled using the cut-and-shifted LJ potential,
\begin{equation}
U_{LJ}^{nb}(r) = 4\epsilon_{ij}[(\frac{\sigma}{r})^{12}-(\frac{\sigma}{r})^6] - 4\epsilon_{ij}[(\frac{\sigma}{r_{cut}})^{12}-(\frac{\sigma}{r_{cut}})^6],
\end{equation}
where the cutoff distance $r_{cut} = 2.5\sigma$ and $\sigma = 1.0$ for all LJ sites. The LJ interaction parameters between all species are set to be $\epsilon_{ij} = 1.0$. Bonding is maintained through a harmonic bond potential
\begin{equation}
U_{LJ}^{b}(r) = \frac{K}{2}(r-\sigma)^2,
\end{equation}
where $K = 2000 (\epsilon/\sigma^2)$. The isothermal-isobaric (NPT) molecular dynamics simulation is used to equilibrate the melt. The system temperature is kept at $T = 0.7$ and the pressure is maintained at $P = 1.0$ using the Nose-Hoover thermostat and barostat. The integration time step for the velocity-Verlet algorithm is set to be $0.002\tau$, where $\tau$ is the unit LJ time.

% Reaction
We then allow the reactions to take place between the chain ends and the crosslinkers. During a crosslinking event, irreversible covalent bonds are formed any time reactive species come within $1.1\sigma$ of each other. Polymer chain ends are only allowed to form one new bond, and crosslinkers are limited to a maximum of four new bonds, following the $A_2 + B_4$ chemistry. The topological defects are naturally emergent from this reactive MD method. \cite{yeMolecular2020} The reaction stage is run for a sufficiently long duration $(100,000 \tau)$ to ensure that the reaction extent is $> 96.0 \%$ and $> 99.5\%$ of crosslinkers are in one system-spanning percolated network.

% Deformation
After the network is formed and equilibrated, we modify the bonding interactions to the breakable quartic potential to allow chain scission for fracture studies. The potential we use is the same as that employed by Ge et al.\cite{geMolecular2013} and Barney et al., \cite{barneyFracture2022}
\begin{equation}
U_Q^b(r)=
K_Q(r-R_c)^3(r-R_c-B)+U_o+4\varepsilon\bigg[\bigg(\frac{\sigma}{r_{cut}}\bigg)^{12}-\bigg(\frac{\sigma}{r_{cut}}\bigg)^6\bigg]+\varepsilon,
\end{equation}
where $K_Q = 2351 (\epsilon/\sigma^4)$, $R_c = 1.5\sigma$, $B = -0.7425\sigma$, and $U_0 = 92.74467\varepsilon$. The network is re-equilibrated prior to deformation under the quartic bond potential. While performing the uniaxial tensile deformation, the simulation box is expanded uniaxially along an axis until all bonds lying along on a plane are broken. The pressure in the other two directions is kept constant while performing the deformation.

\subsection{Isoconfigurational ensemble}
The effect of the network structure on the failure locations in polymer networks is investigated using the isoconfigurational ensemble \cite{widmer-cooperHow2004, colomboMicroscopic2013}. Using the same network configuration, the velocities of all particles are reinitialized in each run prior to deformation. Uniaxial tensile deformations are then performed for all runs with the same direction and strain rate. The isoconfigurational ensemble approach enables us to determine whether the fracture process in polymer networks is purely stochastic or influenced by the underlying network structure. If the failure locations are not structure related, a random set of strands will break in each run. However, if the failure locations are structure related, there will be a significant overlap between the strands broken across multiple runs. When networks fracture, the IDs of broken strands are recorded in each run. The probability of breaking $P_{break}$ of each polymer strand $i$ is defined as
\begin{equation}
P_{i, break} = \frac{n_{i, break}}{n_{runs}}
\end{equation}
where $n_{i, break}$ is the number of times strand $i$ breaks in the isoconfigurational ensemble and $n_{runs} = 10$ for each configuration. For each configuration, the isoconfigurational ensemble is performed in all three directions to ensure that the results are independent of the direction of stretching.

\subsection{Network analysis and GEBC}
Network analysis is performed using the NetworkX package implemented in Python. \cite{SciPyProceedings_11} When constructing the graph based on polymer network structures, crosslinkers of polymer networks are represented as nodes, and polymer chains are represented as edges. Primary loops in polymer networks are represented as self-loops, and secondary and higher order loops are denoted as parallel edges between nodes. Dangling ends are not included in the graphs. The size of graphs constructed in this way depends on the number of crosslinkers and polymer chains. Since the number of monomers $n_{monomer}$ is kept constant across different systems, systems with different strand length $N$ will have graphs of different sizes. The influence of the size of the systems and the graphs on the results will be discussed.

The initial undeformed network structure is represented as an unweighted graph to calculate GEBC for all edges. GEBC of an edge $e$ is defined as the sum of the fraction of all-pairs shortest paths that pass through it \cite{SciPyProceedings_11, brandesFaster2001, brandesVariants2008, newmanNetworks2018},
\begin{equation}
GEBC_{e} = \frac{2}{n(n-1)}\sum_{s,t}\frac{\sigma(s,t|e)}{\sigma(s,t)}
\end{equation}
where $\sigma(s,t)$ is the number of shortest paths between node $s$ and $t$, $\sigma(s,t|e)$ is the number of shortest paths between node $s$ and $t$ that pass through $e$. Each GEBC value is normalized by $2/(n(n-1))$ to restrict the value of GEBC to a range between $0$ and $1$. Additionally, following the approach employed in Berthier et al. \cite{berthierForecasting2019}, we further normalize GEBC by the average value of GEBC of all edges in the network to compare the relative importance of edges to others.

\subsection{Strand orientation}
To investigate the geometric factor behind potential failure locations in polymer networks, the angle between each polymer strand and an axis $\alpha$, $\theta_{\alpha}$ is calculated based on the time-averaged configuration of the equilibrated network. The second Legendre polynomial is used to quantify the orientation of each polymer strand: \cite{gennesPhysics1993}
\begin{equation}
P_{2,\alpha}(cos\theta_{\alpha}) = \frac{1}{2}(3cos^2\theta_{\alpha} - 1)
\end{equation}
the $P_{2,\alpha}$ value is closer to $1$ if the strand is more aligned with the axis $\alpha$ and is closer to $-1/2$ if the strand is more orthogonal to the axis $\alpha$.

\section{Results and discussion}
\subsection{Network structure influences failure location}
\begin{figure}[!ht]
    \centering
    \includegraphics*[width=6.0in]{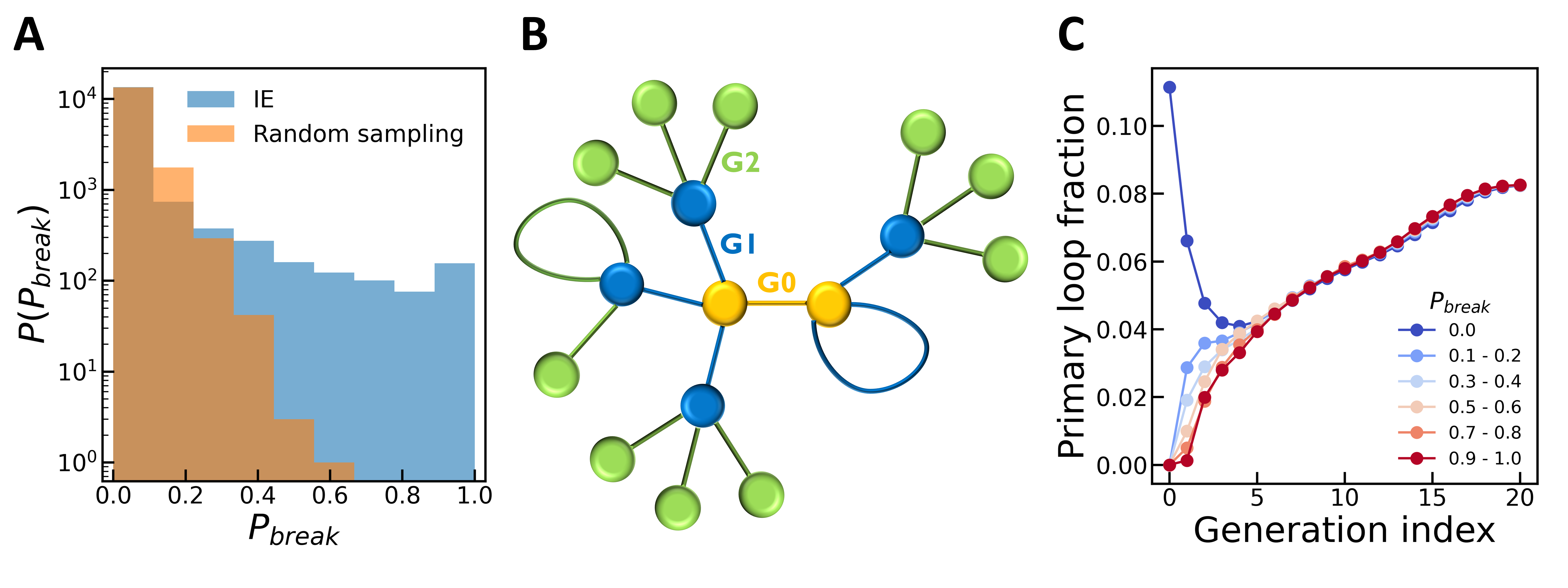}
    \caption{For the $N = 5$ system, (A) distribution of probability of breaking ($P_{break}$) of polymer strands obtained from the isoconfigurational ensemble (blue) and the random sampling (orange); (B) schematic illustration showing the first three generations of the local structure of a polymer strand, strands and crosslinkers at the same generation from the root (G0) strand are labeled with the same color;  (C) primary loop fraction as a function of generation for strands with different probability of breaking in the isoconfigurational ensemble.}
    \label{fgr:p_break}
\end{figure}

We begin by demonstrating at the strands that fail during deformation are not a random subset of polymer strains. The distribution of $P_{break}$ for polymer strands obtained from the isoconfigurational ensemble is compared with a random sampling test to determine whether the fracture of end-linked polymer networks is a purely stochastic process. In each random sampling run, an equal number of broken strands as in each isoconfigurational ensemble run are randomly drawn. Blue histograms in Figure \ref{fgr:p_break}(A) show the results from the isoconfigurational ensemble and orange histograms represent the results from the random sampling. The distribution of $P_{break}$ between these two tests exhibits significant differences, with a distinct subset of polymer strands with very high $P_{break}$ in the isoconfigurational ensemble. This finding suggests that the fracture process in polymer networks is not purely stochastic and is influenced by the network structure.

To explore the relationship between the network structure and the fracture behavior, we start with measuring the defect concentration in the local environment of each polymer strand. As depicted in Figure \ref{fgr:p_break}(B), a polymer strand is chosen as the root (generation 0 strand), and other strands in the network can be located based on their topological distance (generation) from the root strand. For each polymer strand in the network, we can find its local environment and construct a subgraph including all strands within a certain range of generations of that strand. Then we group strands by their $P_{break}$ obtained from the isoconfigurational ensemble and calculate average primary loop fraction as a function of the generation index. The primary loop fraction represents the ratio of the number of primary loops to the total number of polymer strands in each subgraph. Figure \ref{fgr:p_break}(C) illustrates that for strands with high $P_{break}$, their local environment contains fewer primary loops compared to those with low $P_{break}$. The local environment analysis reveals that polymer strands with a relatively defect-free local structure are more susceptible to failure compared to those with a higher concentration defects in their local surroundings.

\subsection{Distribution of GEBC and $P_2$}

\begin{figure}[!ht]
    \centering
    \includegraphics*[width=6.0in]{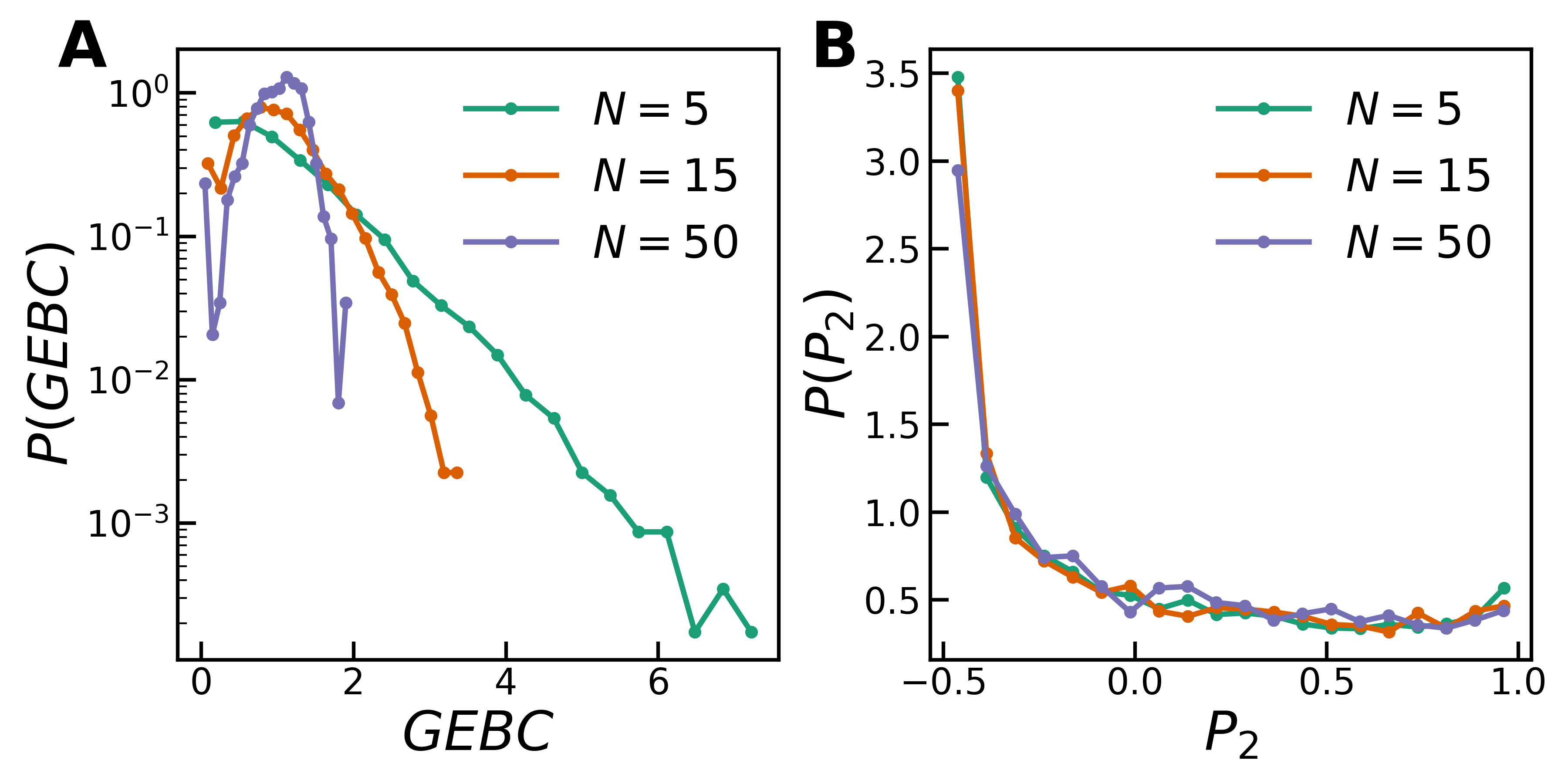}
    \caption{Probability distribution function of (A) GEBC and (B) $P_2$ for the $N = 5, 15$ and $50$ systems. GEBC and $P_2$ are calculated from the initial undeformed state of the network. GEBC of each polymer strand is normalized by the average value of all strands in the network.}
    \label{fgr:dist}
\end{figure}

Mapping polymer networks to graphs and utilizing network analysis tools provide an alternative perspective for probing the structure of these complex disordered systems. The probability distribution functions of GEBC and $P_2$ of polymer strands, calculated from the initial undeformed state of networks, are shown in Figure \ref{fgr:dist}. While the shape of the distribution of $P_2$ is approximately constant across systems with different $N$, the shape of the distribution of GEBC varies. We hypothesize that this variation can be attributed to two reasons. Firstly, networks with higher $N$ at the same polymer volume fraction contain fewer loop defects \cite{wangUniversal2016, zhangPercolation2023}, resulting in a more uniform distribution of GEBC compared to systems with lower $N$ that have a higher defect concentration. Secondly, since the number of polymer beads is kept constant across systems with different $N$, the networks constructed will contain different numbers of polymer strands and crosslinkers, leading to graphs of varying sizes. 

\begin{figure}[!ht]
    \centering
    \includegraphics*[width=6.0in]{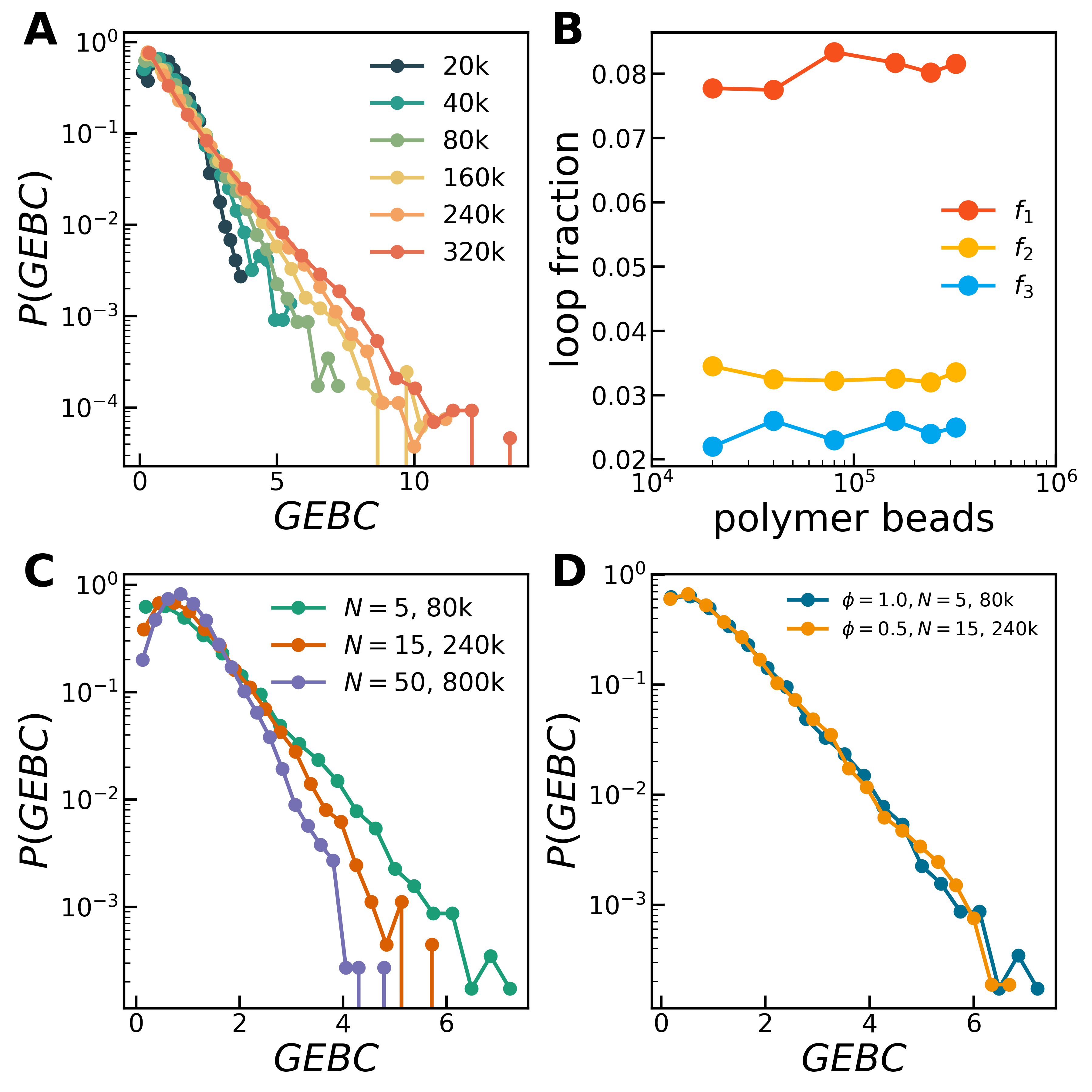}
    \caption{(A) Probability distribution function of GEBC and (B) loop fraction for the $N = 5$ system with various system sizes, ranging from 20,000 to 320,000 polymer beads; (C) Probability distribution function of GEBC for different $N$ systems with the same graph size; (D) Probability distribution function of GEBC for the $\phi=1.0, N = 5$ system with 80,000 polymer beads and $\phi=0.5, N = 15$ system with 240,000 polymer beads. These two systems have approximately the same graph size and the primary loop fraction.}
    \label{fgr:GEBC_dist}
\end{figure}

To test this hypothesis, we construct systems with various strand lengths $N$ and numbers of polymer beads. We first conduct a finite size analysis on the $N = 5$ systems. $N = 5$ systems with polymer beads ranging from $20,000$ to $320,000$ are constructed and their probability distribution functions of GEBC are shown in Figure \ref{fgr:GEBC_dist}(A). The distribution of GEBC becomes wider with increasing the system size. In contrast, the loop fractions are found to be independent of the system size as shown in Figure \ref{fgr:GEBC_dist}(B). Systems with the same graph size but different $N$ are also constructed, and the loop fractions in these systems decrease with strand length $N$. Figure \ref{fgr:GEBC_dist}(C) demonstrates that systems with a higher concentration of topological loop defects exhibit a wider distribution of GEBC. Furthermore, we compare the distribution of GEBC in two systems: a $\phi=1.0, N = 5$ system with $80,000$ polymer beads and a $\phi=0.5, N = 15$ system with $240,000$ polymer beads, where $\phi$ is the polymer volume fraction. Explicit solvent is added to the $\phi=0.5, N = 15$ system to achieve the desired polymer volume fraction. These two systems have approximately the same graph size and loop fraction. Figure \ref{fgr:GEBC_dist}(D) demonstrates that the distribution of GEBC in these two systems is very similar, supporting our hypothesis that the distribution of GEBC is influenced by the system size and the loop fraction. Overall, the distribution of GEBC in all systems exhibits a shape resembling the Gamma function, indicating that a small subset of polymer strands within the networks have GEBC values several times higher than the system average.

\subsection{GEBC and $P_2$ are useful in identifying the failure location}

\begin{figure}[!ht]
    \centering
    \includegraphics*[width=6.0in]{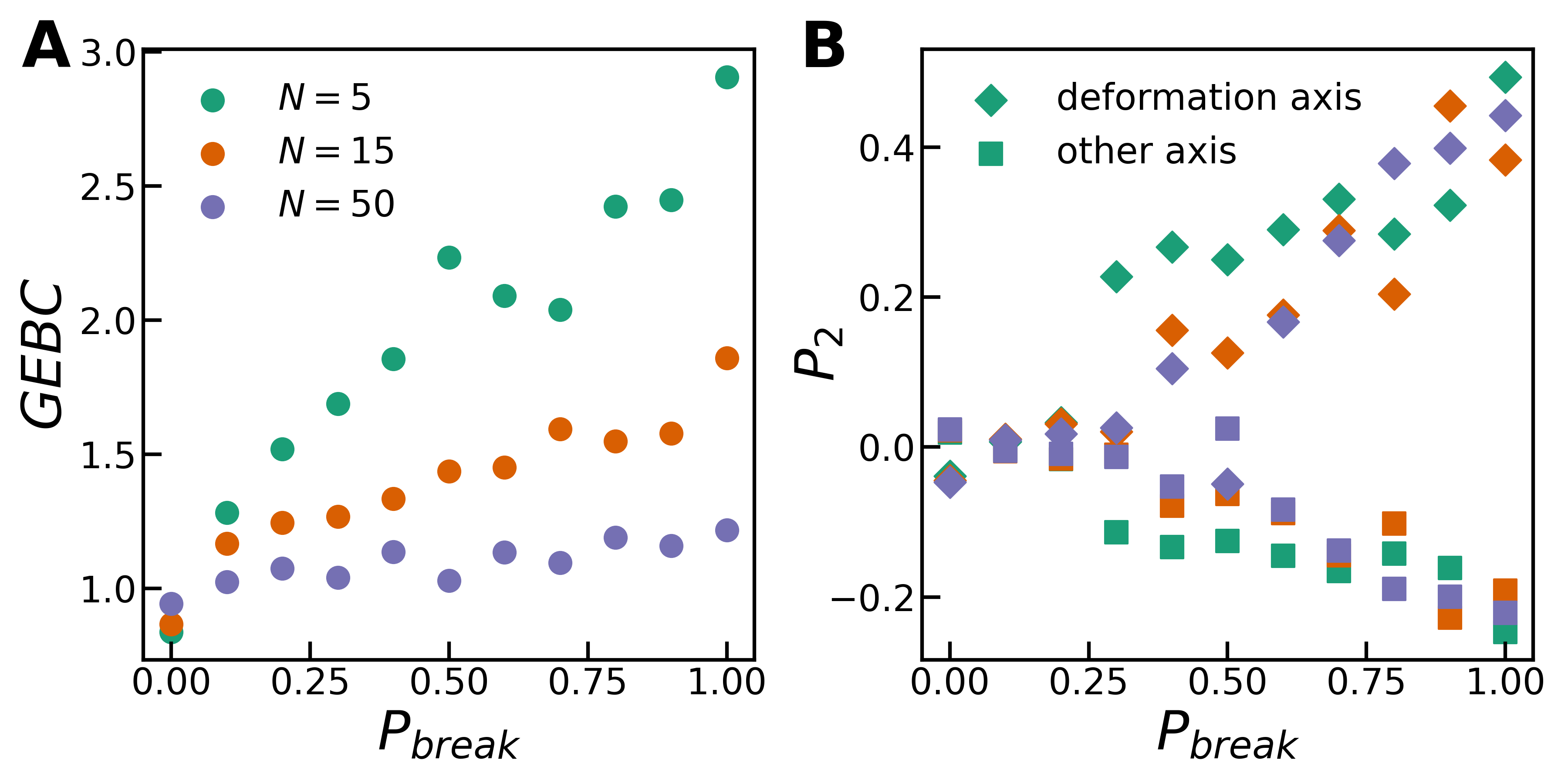}
    \caption{Average (A) GEBC and (B) $P_2$ for polymer strands with the same $P_{break}$ for the $N = 5, 15$ and $50$ network systems.}
    \label{fgr:GEBC_P2_p_break}
\end{figure}

We proceed to group strands with the same $P_{break}$ from the isoconfigurational ensemble together to investigate the underlying topological and geometric factor of the network structure influencing the potential failure locations. Figure \ref{fgr:GEBC_P2_p_break}(A) demonstrates a consistent positive correlation between the average GEBC of each group and $P_{break}$ across all systems, regardless of the shape of GEBC distribution in these systems. This indicates that in the end-linked network systems, on average, strands with higher GEBC have a higher probability of breaking. Moreover, Figure \ref{fgr:GEBC_P2_p_break}(B) illustrates a positive correlation between average $P_2$ on the deformation axis and $P_{break}$, as well as a negative correlation between average $P_2$ in the orthogonal directions and $P_{break}$. This finding suggests that strands that are more aligned with the deformation axis have a higher probability of breaking.

\begin{figure}[!ht]
    \centering
    \includegraphics*[width=6.0in]{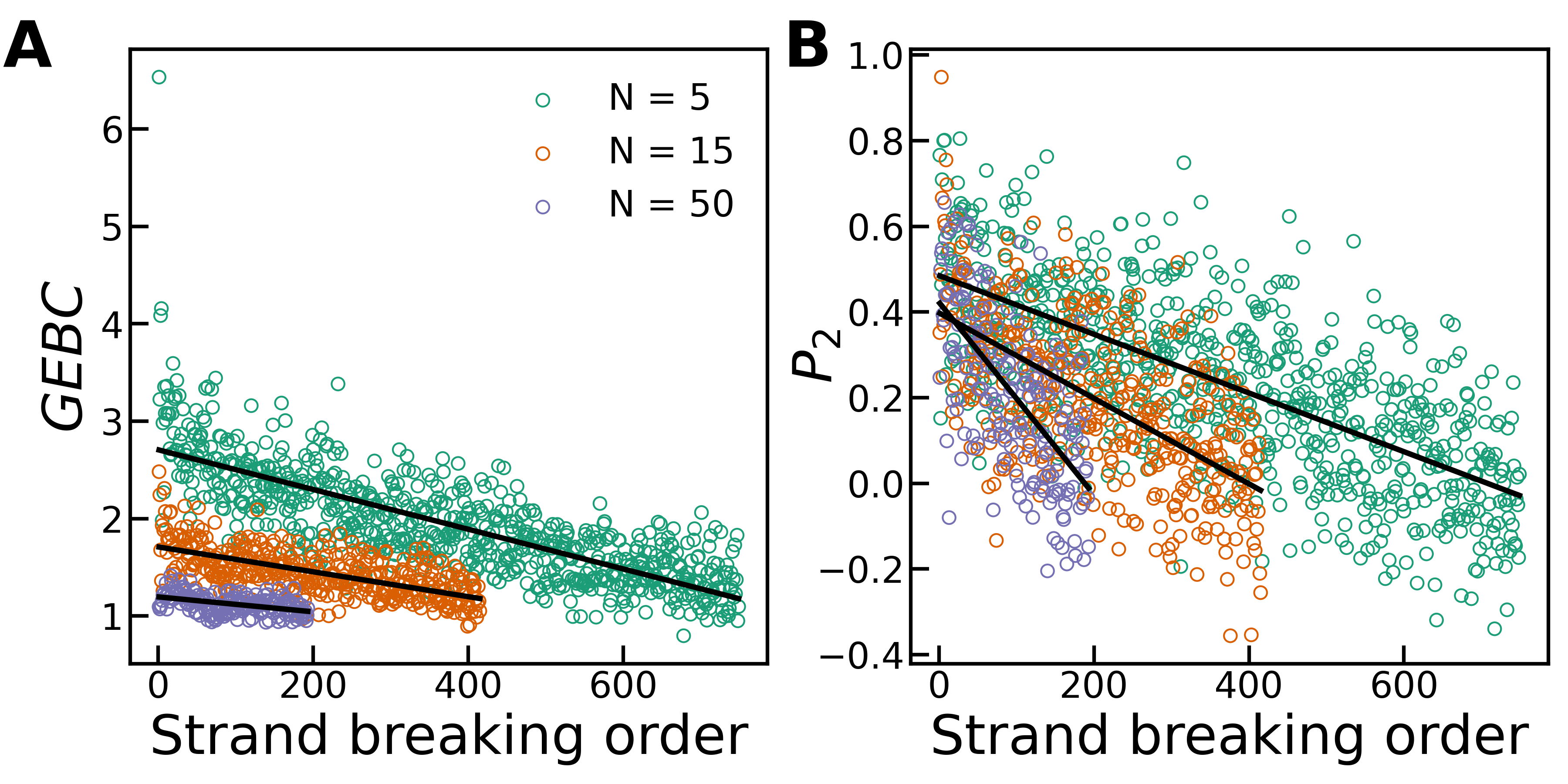}
    \caption{(A) GEBC and (B) $P_2$ as a function of strand breaking order for the $N = 5, 15$ and $50$ systems. Black lines are linear fittings of each dataset.}
    \label{fgr:order}
\end{figure}

GEBC and $P_2$ are also calculated as a function of the order in which the strands break during the deformation. Since the order in which the strands break can vary between each realization in the isoconfigurational ensemble, the average GEBC and $P_2$ is calculated independently in each trajectory and plotted in Figure \ref{fgr:order}. Both GEBC and $P_2$ are highest for the first several chains that break then appear to decrease linearly as the deformation proceeds towards failiure.
%At each order of strand breaking, the average GEBC and $P_2$ of polymer strands are calculated across all isoconfigurational ensemble runs. 
In Figure \ref{fgr:order}, the black lines in the plots represent linear fittings of these data points. The linear trend suggests that strand breaking events initiate from strands with higher GEBC and smaller angles between the polymer strands and the deformation axis. We note that when the network approaches failure, GEBC approaches a value close to $1$ and $P_2$ decreases to approximately $0.0$, which implies that the predicting power of these two factors diminishes as the network approaches the fracture point, but these measures are important in the initial stages of failure. These results also show the importance of topological and geometric features of the networks beyond the more common network defects such as loops and dangling ends.

\section{Conclusions}
 This work provides valuable insights into the interplay between network topology, strand orientation relative to the direction of strain, and the fracture behavior of model end-linked polymer networks. Through a combination of molecular simulations and network analysis techniques, we reveal that polymer strands with relatively defect-free local structures, high GEBC values and greater alignment with the deformation axis are more susceptible to failure during uniaxial tensile deformation. Additionally, strand breaking events initiate from polymer strands with higher GEBC values and smaller angles between the strand and the deformation axis. The effectiveness of GEBC in identifying the failure locations is consistent with recent observations in other types of network materials, including granular networks \cite{berthierForecasting2019} and colloidal gels \cite{mangalTopological2023}, highlighting the potential of network analysis tools in finding universal properties across different network systems. Future work will explore the predictive capabilities of GEBC and strand orientation in other types of reaction schemes and deformations, as well as further investigate other network analysis techniques such as community detection \cite{fazelpourCommunity2023}. These findings also offer a novel perspective to understand the influence of topological defects in polymer networks and suggest an route to inversely design polymer network structures to control the location of fracture and achieve desired fracture properties.

\begin{acknowledgement}
The authors acknowledge support from the Office of Naval Research via ONR-N00014-17-1-2056. Computational resources were made available through XSEDE Award TG-DMR150034 and Access Award CHM230005. 
\end{acknowledgement}

% \begin{suppinfo}

% This will usually read something like: ``Experimental procedures and
% characterization data for all new compounds. The class will
% automatically add a sentence pointing to the information on-line:

% \end{suppinfo}

%\bibliography{output}

\begin{mcitethebibliography}{55}
\providecommand*\natexlab[1]{#1}
\providecommand*\mciteSetBstSublistMode[1]{}
\providecommand*\mciteSetBstMaxWidthForm[2]{}
\providecommand*\mciteBstWouldAddEndPuncttrue
  {\def\EndOfBibitem{\unskip.}}
\providecommand*\mciteBstWouldAddEndPunctfalse
  {\let\EndOfBibitem\relax}
\providecommand*\mciteSetBstMidEndSepPunct[3]{}
\providecommand*\mciteSetBstSublistLabelBeginEnd[3]{}
\providecommand*\EndOfBibitem{}
\mciteSetBstSublistMode{f}
\mciteSetBstMaxWidthForm{subitem}{(\alph{mcitesubitemcount})}
\mciteSetBstSublistLabelBeginEnd
  {\mcitemaxwidthsubitemform\space}
  {\relax}
  {\relax}

\bibitem[Rogers \latin{et~al.}(2010)Rogers, Someya, and
  Huang]{rogersMaterials2010}
Rogers,~J.~A.; Someya,~T.; Huang,~Y. Materials and Mechanics for Stretchable
  Electronics. \emph{Science} \textbf{2010}, \emph{327}, 1603--1607\relax
\mciteBstWouldAddEndPuncttrue
\mciteSetBstMidEndSepPunct{\mcitedefaultmidpunct}
{\mcitedefaultendpunct}{\mcitedefaultseppunct}\relax
\EndOfBibitem
\bibitem[Ionov(2014)]{ionovHydrogelbased2014}
Ionov,~L. Hydrogel-Based Actuators: Possibilities and Limitations.
  \emph{Materials Today} \textbf{2014}, \emph{17}, 494--503\relax
\mciteBstWouldAddEndPuncttrue
\mciteSetBstMidEndSepPunct{\mcitedefaultmidpunct}
{\mcitedefaultendpunct}{\mcitedefaultseppunct}\relax
\EndOfBibitem
\bibitem[Li and Mooney(2016)Li, and Mooney]{liDesigning2016}
Li,~J.; Mooney,~D.~J. Designing Hydrogels for Controlled Drug Delivery.
  \emph{Nature Reviews Materials} \textbf{2016}, \emph{1}, 1--17\relax
\mciteBstWouldAddEndPuncttrue
\mciteSetBstMidEndSepPunct{\mcitedefaultmidpunct}
{\mcitedefaultendpunct}{\mcitedefaultseppunct}\relax
\EndOfBibitem
\bibitem[Gu \latin{et~al.}(2020)Gu, Zhao, and Johnson]{guPolymer2020}
Gu,~Y.; Zhao,~J.; Johnson,~J.~A. Polymer {{Networks}}: {{From Plastics}} and
  {{Gels}} to {{Porous Frameworks}}. \emph{Angewandte Chemie International
  Edition} \textbf{2020}, \emph{59}, 5022--5049\relax
\mciteBstWouldAddEndPuncttrue
\mciteSetBstMidEndSepPunct{\mcitedefaultmidpunct}
{\mcitedefaultendpunct}{\mcitedefaultseppunct}\relax
\EndOfBibitem
\bibitem[Danielsen \latin{et~al.}(2021)Danielsen, Beech, Wang, {El-Zaatari},
  Wang, Sapir, Ouchi, Wang, Johnson, Hu, Lundberg, Stoychev, Craig, Johnson,
  Kalow, Olsen, and Rubinstein]{danielsenMolecular2021}
Danielsen,~S. P.~O. \latin{et~al.}  Molecular {{Characterization}} of {{Polymer
  Networks}}. \emph{Chemical Reviews} \textbf{2021}, \emph{121},
  5042--5092\relax
\mciteBstWouldAddEndPuncttrue
\mciteSetBstMidEndSepPunct{\mcitedefaultmidpunct}
{\mcitedefaultendpunct}{\mcitedefaultseppunct}\relax
\EndOfBibitem
\bibitem[Zhao \latin{et~al.}(2021)Zhao, Chen, Yuk, Lin, Liu, and
  Parada]{zhaoSoft2021}
Zhao,~X.; Chen,~X.; Yuk,~H.; Lin,~S.; Liu,~X.; Parada,~G. Soft {{Materials}} by
  {{Design}}: {{Unconventional Polymer Networks Give Extreme Properties}}.
  \emph{Chemical Reviews} \textbf{2021}, \emph{121}, 4309--4372\relax
\mciteBstWouldAddEndPuncttrue
\mciteSetBstMidEndSepPunct{\mcitedefaultmidpunct}
{\mcitedefaultendpunct}{\mcitedefaultseppunct}\relax
\EndOfBibitem
\bibitem[Zhou \latin{et~al.}(2023)Zhou, Yuk, Hu, Wu, Tian, Roh, Shen, Gu, Xu,
  Lu, and Zhao]{zhou3D2023}
Zhou,~T.; Yuk,~H.; Hu,~F.; Wu,~J.; Tian,~F.; Roh,~H.; Shen,~Z.; Gu,~G.; Xu,~J.;
  Lu,~B.; Zhao,~X. {{3D}} Printable High-Performance Conducting Polymer
  Hydrogel for All-Hydrogel Bioelectronic Interfaces. \emph{Nature Materials}
  \textbf{2023}, 1--8\relax
\mciteBstWouldAddEndPuncttrue
\mciteSetBstMidEndSepPunct{\mcitedefaultmidpunct}
{\mcitedefaultendpunct}{\mcitedefaultseppunct}\relax
\EndOfBibitem
\bibitem[Tauber \latin{et~al.}(2022)Tauber, {van der Gucht}, and
  Dussi]{tauberStretchy2022}
Tauber,~J.; {van der Gucht},~J.; Dussi,~S. Stretchy and Disordered: {{Toward}}
  Understanding Fracture in Soft Network Materials via Mesoscopic Computer
  Simulations. \emph{The Journal of Chemical Physics} \textbf{2022},
  \emph{156}, 160901\relax
\mciteBstWouldAddEndPuncttrue
\mciteSetBstMidEndSepPunct{\mcitedefaultmidpunct}
{\mcitedefaultendpunct}{\mcitedefaultseppunct}\relax
\EndOfBibitem
\bibitem[Creton(2017)]{creton50th2017}
Creton,~C. 50th {{Anniversary Perspective}}: {{Networks}} and {{Gels}}:
  {{Soft}} but {{Dynamic}} and {{Tough}}. \emph{Macromolecules} \textbf{2017},
  \emph{50}, 8297--8316\relax
\mciteBstWouldAddEndPuncttrue
\mciteSetBstMidEndSepPunct{\mcitedefaultmidpunct}
{\mcitedefaultendpunct}{\mcitedefaultseppunct}\relax
\EndOfBibitem
\bibitem[Zhou \latin{et~al.}(2012)Zhou, Woo, Cok, Wang, Olsen, and
  Johnson]{zhouCounting2012}
Zhou,~H.; Woo,~J.; Cok,~A.~M.; Wang,~M.; Olsen,~B.~D.; Johnson,~J.~A. Counting
  Primary Loops in Polymer Gels. \emph{Proceedings of the National Academy of
  Sciences} \textbf{2012}, \emph{109}, 19119--19124\relax
\mciteBstWouldAddEndPuncttrue
\mciteSetBstMidEndSepPunct{\mcitedefaultmidpunct}
{\mcitedefaultendpunct}{\mcitedefaultseppunct}\relax
\EndOfBibitem
\bibitem[Zhong \latin{et~al.}(2016)Zhong, Wang, Kawamoto, Olsen, and
  Johnson]{zhongQuantifying2016}
Zhong,~M.; Wang,~R.; Kawamoto,~K.; Olsen,~B.~D.; Johnson,~J.~A. Quantifying the
  Impact of Molecular Defects on Polymer Network Elasticity. \emph{Science}
  \textbf{2016}, \emph{353}, 1264--1268\relax
\mciteBstWouldAddEndPuncttrue
\mciteSetBstMidEndSepPunct{\mcitedefaultmidpunct}
{\mcitedefaultendpunct}{\mcitedefaultseppunct}\relax
\EndOfBibitem
\bibitem[Wang \latin{et~al.}(2016)Wang, {Alexander-Katz}, Johnson, and
  Olsen]{wangUniversal2016}
Wang,~R.; {Alexander-Katz},~A.; Johnson,~J.~A.; Olsen,~B.~D. Universal {{Cyclic
  Topology}} in {{Polymer Networks}}. \emph{Physical Review Letters}
  \textbf{2016}, \emph{116}, 188302\relax
\mciteBstWouldAddEndPuncttrue
\mciteSetBstMidEndSepPunct{\mcitedefaultmidpunct}
{\mcitedefaultendpunct}{\mcitedefaultseppunct}\relax
\EndOfBibitem
\bibitem[Lang(2018)]{langElasticity2018}
Lang,~M. Elasticity of {{Phantom Model Networks}} with {{Cyclic Defects}}.
  \emph{ACS Macro Letters} \textbf{2018}, \emph{7}, 536--539\relax
\mciteBstWouldAddEndPuncttrue
\mciteSetBstMidEndSepPunct{\mcitedefaultmidpunct}
{\mcitedefaultendpunct}{\mcitedefaultseppunct}\relax
\EndOfBibitem
\bibitem[Panyukov(2019)]{panyukovLoops2019}
Panyukov,~S. Loops in {{Polymer Networks}}. \emph{Macromolecules}
  \textbf{2019}, \emph{52}, 4145--4153\relax
\mciteBstWouldAddEndPuncttrue
\mciteSetBstMidEndSepPunct{\mcitedefaultmidpunct}
{\mcitedefaultendpunct}{\mcitedefaultseppunct}\relax
\EndOfBibitem
\bibitem[Lin \latin{et~al.}(2019)Lin, Wang, Johnson, and
  Olsen]{linRevisiting2019}
Lin,~T.~S.; Wang,~R.; Johnson,~J.~A.; Olsen,~B.~D. Revisiting the {{Elasticity
  Theory}} for {{Real Gaussian Phantom Networks}}. \emph{Macromolecules}
  \textbf{2019}, \emph{52}, 1685--1694\relax
\mciteBstWouldAddEndPuncttrue
\mciteSetBstMidEndSepPunct{\mcitedefaultmidpunct}
{\mcitedefaultendpunct}{\mcitedefaultseppunct}\relax
\EndOfBibitem
\bibitem[Lin and Zhao(2020)Lin, and Zhao]{linFracture2020}
Lin,~S.; Zhao,~X. Fracture of Polymer Networks with Diverse Topological
  Defects. \emph{Physical Review E} \textbf{2020}, \emph{102}, 052503\relax
\mciteBstWouldAddEndPuncttrue
\mciteSetBstMidEndSepPunct{\mcitedefaultmidpunct}
{\mcitedefaultendpunct}{\mcitedefaultseppunct}\relax
\EndOfBibitem
\bibitem[Arora \latin{et~al.}(2020)Arora, Lin, Beech, Mochigase, Wang, and
  Olsen]{aroraFracture2020}
Arora,~A.; Lin,~T.~S.; Beech,~H.~K.; Mochigase,~H.; Wang,~R.; Olsen,~B.~D.
  Fracture of {{Polymer Networks Containing Topological Defects}}.
  \emph{Macromolecules} \textbf{2020}, \emph{53}, 7346--7355\relax
\mciteBstWouldAddEndPuncttrue
\mciteSetBstMidEndSepPunct{\mcitedefaultmidpunct}
{\mcitedefaultendpunct}{\mcitedefaultseppunct}\relax
\EndOfBibitem
\bibitem[Arora \latin{et~al.}(2022)Arora, Lin, and
  Olsen]{aroraCoarseGrained2022}
Arora,~A.; Lin,~T.~S.; Olsen,~B.~D. Coarse-{{Grained Simulations}} for
  {{Fracture}} of {{Polymer Networks}}: {{Stress Versus Topological
  Inhomogeneities}}. \emph{Macromolecules} \textbf{2022}, \emph{55},
  4--14\relax
\mciteBstWouldAddEndPuncttrue
\mciteSetBstMidEndSepPunct{\mcitedefaultmidpunct}
{\mcitedefaultendpunct}{\mcitedefaultseppunct}\relax
\EndOfBibitem
\bibitem[Barney \latin{et~al.}(2022)Barney, Ye, Sacligil, McLeod, Zhang, Tew,
  Riggleman, and Crosby]{barneyFracture2022}
Barney,~C.~W.; Ye,~Z.; Sacligil,~I.; McLeod,~K.~R.; Zhang,~H.; Tew,~G.~N.;
  Riggleman,~R.~A.; Crosby,~A.~J. Fracture of Model End-Linked Networks.
  \emph{Proceedings of the National Academy of Sciences} \textbf{2022},
  \emph{119}, e2112389119\relax
\mciteBstWouldAddEndPuncttrue
\mciteSetBstMidEndSepPunct{\mcitedefaultmidpunct}
{\mcitedefaultendpunct}{\mcitedefaultseppunct}\relax
\EndOfBibitem
\bibitem[Rubinstein and Colby(2003)Rubinstein, and
  Colby]{rubinsteinPolymer2003}
Rubinstein,~M.; Colby,~R.~H. In \emph{Polymer {{Physics}}}; Colby,~R.~H., Ed.;
  {OUP Oxford}: {Oxford}, 2003\relax
\mciteBstWouldAddEndPuncttrue
\mciteSetBstMidEndSepPunct{\mcitedefaultmidpunct}
{\mcitedefaultendpunct}{\mcitedefaultseppunct}\relax
\EndOfBibitem
\bibitem[Lake and Thomas(1967)Lake, and Thomas]{lakeStrength1967}
Lake,~G.~J.; Thomas,~A.~G. The Strength of Highly Elastic Materials.
  \emph{Proceedings of the Royal Society of London. Series A. Mathematical and
  Physical Sciences} \textbf{1967}, \emph{300}, 108--119\relax
\mciteBstWouldAddEndPuncttrue
\mciteSetBstMidEndSepPunct{\mcitedefaultmidpunct}
{\mcitedefaultendpunct}{\mcitedefaultseppunct}\relax
\EndOfBibitem
\bibitem[Newman(2018)]{newmanNetworks2018}
Newman,~M. \emph{Networks}; {Oxford University Press}, 2018\relax
\mciteBstWouldAddEndPuncttrue
\mciteSetBstMidEndSepPunct{\mcitedefaultmidpunct}
{\mcitedefaultendpunct}{\mcitedefaultseppunct}\relax
\EndOfBibitem
\bibitem[Bassett \latin{et~al.}(2012)Bassett, Owens, Daniels, and
  Porter]{bassettInfluence2012}
Bassett,~D.~S.; Owens,~E.~T.; Daniels,~K.~E.; Porter,~M.~A. Influence of
  Network Topology on Sound Propagation in Granular Materials. \emph{Physical
  Review E} \textbf{2012}, \emph{86}, 041306\relax
\mciteBstWouldAddEndPuncttrue
\mciteSetBstMidEndSepPunct{\mcitedefaultmidpunct}
{\mcitedefaultendpunct}{\mcitedefaultseppunct}\relax
\EndOfBibitem
\bibitem[Papadopoulos \latin{et~al.}(2018)Papadopoulos, Porter, Daniels, and
  Bassett]{papadopoulosNetwork2018}
Papadopoulos,~L.; Porter,~M.~A.; Daniels,~K.~E.; Bassett,~D.~S. Network
  Analysis of Particles and Grains. \emph{Journal of Complex Networks}
  \textbf{2018}, \emph{6}, 485--565\relax
\mciteBstWouldAddEndPuncttrue
\mciteSetBstMidEndSepPunct{\mcitedefaultmidpunct}
{\mcitedefaultendpunct}{\mcitedefaultseppunct}\relax
\EndOfBibitem
\bibitem[Berthier \latin{et~al.}(2019)Berthier, Porter, and
  Daniels]{berthierForecasting2019}
Berthier,~E.; Porter,~M.~A.; Daniels,~K.~E. Forecasting Failure Locations in
  2-Dimensional Disordered Lattices. \emph{Proceedings of the National Academy
  of Sciences} \textbf{2019}, \emph{116}, 16742--16749\relax
\mciteBstWouldAddEndPuncttrue
\mciteSetBstMidEndSepPunct{\mcitedefaultmidpunct}
{\mcitedefaultendpunct}{\mcitedefaultseppunct}\relax
\EndOfBibitem
\bibitem[Kollmer and Daniels(2019)Kollmer, and Daniels]{kollmerBetweenness2019}
Kollmer,~J.~E.; Daniels,~K.~E. Betweenness Centrality as Predictor for Forces
  in Granular Packings. \emph{Soft Matter} \textbf{2019}, \emph{15},
  1793--1798\relax
\mciteBstWouldAddEndPuncttrue
\mciteSetBstMidEndSepPunct{\mcitedefaultmidpunct}
{\mcitedefaultendpunct}{\mcitedefaultseppunct}\relax
\EndOfBibitem
\bibitem[Pournajar \latin{et~al.}(2022)Pournajar, Zaiser, and
  Moretti]{pournajarEdge2022}
Pournajar,~M.; Zaiser,~M.; Moretti,~P. Edge Betweenness Centrality as a Failure
  Predictor in Network Models of Structurally Disordered Materials.
  \emph{Scientific Reports} \textbf{2022}, \emph{12}, 11814\relax
\mciteBstWouldAddEndPuncttrue
\mciteSetBstMidEndSepPunct{\mcitedefaultmidpunct}
{\mcitedefaultendpunct}{\mcitedefaultseppunct}\relax
\EndOfBibitem
\bibitem[Nabizadeh \latin{et~al.}(2022)Nabizadeh, Singh, and
  Jamali]{nabizadehStructure2022a}
Nabizadeh,~M.; Singh,~A.; Jamali,~S. Structure and {{Dynamics}} of {{Force
  Clusters}} and {{Networks}} in {{Shear Thickening Suspensions}}.
  \emph{Physical Review Letters} \textbf{2022}, \emph{129}, 068001\relax
\mciteBstWouldAddEndPuncttrue
\mciteSetBstMidEndSepPunct{\mcitedefaultmidpunct}
{\mcitedefaultendpunct}{\mcitedefaultseppunct}\relax
\EndOfBibitem
\bibitem[Mangal \latin{et~al.}(2023)Mangal, Nabizadeh, and
  Jamali]{mangalTopological2023}
Mangal,~D.; Nabizadeh,~M.; Jamali,~S. Topological Origins of Yielding in
  Short-Ranged Weakly Attractive Colloidal Gels. \emph{The Journal of Chemical
  Physics} \textbf{2023}, \emph{158}, 014903\relax
\mciteBstWouldAddEndPuncttrue
\mciteSetBstMidEndSepPunct{\mcitedefaultmidpunct}
{\mcitedefaultendpunct}{\mcitedefaultseppunct}\relax
\EndOfBibitem
\bibitem[Fazelpour \latin{et~al.}(2023)Fazelpour, Desai, and
  Daniels]{fazelpourCommunity2023}
Fazelpour,~F.; Desai,~V.~D.; Daniels,~K.~E. Community Detection Forecasts
  Material Failure in a Sheared Granular Material. 2023\relax
\mciteBstWouldAddEndPuncttrue
\mciteSetBstMidEndSepPunct{\mcitedefaultmidpunct}
{\mcitedefaultendpunct}{\mcitedefaultseppunct}\relax
\EndOfBibitem
\bibitem[Moretti and Zaiser(2019)Moretti, and Zaiser]{morettiNetwork2019a}
Moretti,~P.; Zaiser,~M. Network Analysis Predicts Failure of Materials and
  Structures. \emph{Proceedings of the National Academy of Sciences}
  \textbf{2019}, \emph{116}, 16666--16668\relax
\mciteBstWouldAddEndPuncttrue
\mciteSetBstMidEndSepPunct{\mcitedefaultmidpunct}
{\mcitedefaultendpunct}{\mcitedefaultseppunct}\relax
\EndOfBibitem
\bibitem[Freeman(1977)]{freemanSet1977}
Freeman,~L.~C. A {{Set}} of {{Measures}} of {{Centrality Based}} on
  {{Betweenness}}. \emph{Sociometry} \textbf{1977}, \emph{40}, 35--41\relax
\mciteBstWouldAddEndPuncttrue
\mciteSetBstMidEndSepPunct{\mcitedefaultmidpunct}
{\mcitedefaultendpunct}{\mcitedefaultseppunct}\relax
\EndOfBibitem
\bibitem[Grest \latin{et~al.}(1992)Grest, Kremer, and
  Duering]{grestKinetics1992}
Grest,~G.~S.; Kremer,~K.; Duering,~E.~R. Kinetics of {{End Crosslinking}} in
  {{Dense Polymer Melts}}. \emph{Europhysics Letters} \textbf{1992}, \emph{19},
  195\relax
\mciteBstWouldAddEndPuncttrue
\mciteSetBstMidEndSepPunct{\mcitedefaultmidpunct}
{\mcitedefaultendpunct}{\mcitedefaultseppunct}\relax
\EndOfBibitem
\bibitem[Duering \latin{et~al.}(1994)Duering, Kremer, and
  Grest]{dueringStructure1994}
Duering,~E.~R.; Kremer,~K.; Grest,~G.~S. Structure and Relaxation of End-linked
  Polymer Networks. \emph{The Journal of Chemical Physics} \textbf{1994},
  \emph{101}, 8169--8192\relax
\mciteBstWouldAddEndPuncttrue
\mciteSetBstMidEndSepPunct{\mcitedefaultmidpunct}
{\mcitedefaultendpunct}{\mcitedefaultseppunct}\relax
\EndOfBibitem
\bibitem[Nowak and Escobedo(2016)Nowak, and Escobedo]{nowakTuning2016}
Nowak,~C.; Escobedo,~F.~A. Tuning the {{Sawtooth Tensile Response}} and
  {{Toughness}} of {{Multiblock Copolymer Diamond Networks}}.
  \emph{Macromolecules} \textbf{2016}, \emph{49}, 6711--6721\relax
\mciteBstWouldAddEndPuncttrue
\mciteSetBstMidEndSepPunct{\mcitedefaultmidpunct}
{\mcitedefaultendpunct}{\mcitedefaultseppunct}\relax
\EndOfBibitem
\bibitem[Nowak and Escobedo(2017)Nowak, and Escobedo]{nowakOptimizing2017}
Nowak,~C.; Escobedo,~F.~A. Optimizing the Network Topology of Block Copolymer
  Liquid Crystal Elastomers for Enhanced Extensibility and Toughness.
  \emph{Physical Review Materials} \textbf{2017}, \emph{1}, 035601\relax
\mciteBstWouldAddEndPuncttrue
\mciteSetBstMidEndSepPunct{\mcitedefaultmidpunct}
{\mcitedefaultendpunct}{\mcitedefaultseppunct}\relax
\EndOfBibitem
\bibitem[Gusev(2019)]{gusevNumerical2019}
Gusev,~A.~A. Numerical {{Estimates}} of the {{Topological Effects}} in the
  {{Elasticity}} of {{Gaussian Polymer Networks}} and {{Their Exact Theoretical
  Description}}. \emph{Macromolecules} \textbf{2019}, \emph{52},
  3244--3251\relax
\mciteBstWouldAddEndPuncttrue
\mciteSetBstMidEndSepPunct{\mcitedefaultmidpunct}
{\mcitedefaultendpunct}{\mcitedefaultseppunct}\relax
\EndOfBibitem
\bibitem[Lang and M{\"u}ller(2020)Lang, and M{\"u}ller]{langAnalysis2020a}
Lang,~M.; M{\"u}ller,~T. Analysis of the {{Gel Point}} of {{Polymer Model
  Networks}} by {{Computer Simulations}}. \emph{Macromolecules} \textbf{2020},
  \emph{53}, 498--512\relax
\mciteBstWouldAddEndPuncttrue
\mciteSetBstMidEndSepPunct{\mcitedefaultmidpunct}
{\mcitedefaultendpunct}{\mcitedefaultseppunct}\relax
\EndOfBibitem
\bibitem[Ye and Riggleman(2020)Ye, and Riggleman]{yeMolecular2020}
Ye,~Z.; Riggleman,~R.~A. Molecular {{View}} of {{Cavitation}} in
  {{Model-Solvated Polymer Networks}}. \emph{Macromolecules} \textbf{2020},
  \emph{53}, 7825--7834\relax
\mciteBstWouldAddEndPuncttrue
\mciteSetBstMidEndSepPunct{\mcitedefaultmidpunct}
{\mcitedefaultendpunct}{\mcitedefaultseppunct}\relax
\EndOfBibitem
\bibitem[Zhang and Riggleman(2023)Zhang, and Riggleman]{zhangPercolation2023}
Zhang,~H.; Riggleman,~R.~A. Percolation of Co-Continuous Domains in Tapered
  Copolymer Networks. \emph{Molecular Systems Design \& Engineering}
  \textbf{2023}, \emph{8}, 115--122\relax
\mciteBstWouldAddEndPuncttrue
\mciteSetBstMidEndSepPunct{\mcitedefaultmidpunct}
{\mcitedefaultendpunct}{\mcitedefaultseppunct}\relax
\EndOfBibitem
\bibitem[Masubuchi \latin{et~al.}(2023)Masubuchi, Doi, Ishida, Sakumichi,
  Sakai, Mayumi, and Uneyama]{masubuchiPhantom2023}
Masubuchi,~Y.; Doi,~Y.; Ishida,~T.; Sakumichi,~N.; Sakai,~T.; Mayumi,~K.;
  Uneyama,~T. Phantom {{Chain Simulations}} for the {{Fracture}} of
  {{Energy-Minimized Tetra-}} and {{Tri-Branched Networks}}.
  \emph{Macromolecules} \textbf{2023}, \emph{56}, 2217--2223\relax
\mciteBstWouldAddEndPuncttrue
\mciteSetBstMidEndSepPunct{\mcitedefaultmidpunct}
{\mcitedefaultendpunct}{\mcitedefaultseppunct}\relax
\EndOfBibitem
\bibitem[Sorichetti \latin{et~al.}(2023)Sorichetti, Ninarello, {Ruiz-Franco},
  Hugouvieux, Zaccarelli, Micheletti, Kob, and
  Rovigatti]{sorichettiStructure2023}
Sorichetti,~V.; Ninarello,~A.; {Ruiz-Franco},~J.; Hugouvieux,~V.;
  Zaccarelli,~E.; Micheletti,~C.; Kob,~W.; Rovigatti,~L. Structure and
  Elasticity of Model Disordered, Polydisperse, and Defect-Free Polymer
  Networks. \emph{The Journal of Chemical Physics} \textbf{2023}, \emph{158},
  074905\relax
\mciteBstWouldAddEndPuncttrue
\mciteSetBstMidEndSepPunct{\mcitedefaultmidpunct}
{\mcitedefaultendpunct}{\mcitedefaultseppunct}\relax
\EndOfBibitem
\bibitem[Plimpton(1995)]{plimptonFast1995}
Plimpton,~S. Fast {{Parallel Algorithms}} for {{Short-Range Molecular
  Dynamics}}. \emph{Journal of Computational Physics} \textbf{1995},
  \emph{117}, 1--19\relax
\mciteBstWouldAddEndPuncttrue
\mciteSetBstMidEndSepPunct{\mcitedefaultmidpunct}
{\mcitedefaultendpunct}{\mcitedefaultseppunct}\relax
\EndOfBibitem
\bibitem[Thompson \latin{et~al.}(2022)Thompson, Aktulga, Berger, Bolintineanu,
  Brown, Crozier, {in 't Veld}, Kohlmeyer, Moore, Nguyen, Shan, Stevens,
  Tranchida, Trott, and Plimpton]{thompsonLAMMPS2022}
Thompson,~A.~P.; Aktulga,~H.~M.; Berger,~R.; Bolintineanu,~D.~S.; Brown,~W.~M.;
  Crozier,~P.~S.; {in 't Veld},~P.~J.; Kohlmeyer,~A.; Moore,~S.~G.;
  Nguyen,~T.~D.; Shan,~R.; Stevens,~M.~J.; Tranchida,~J.; Trott,~C.;
  Plimpton,~S.~J. {{LAMMPS}} - a Flexible Simulation Tool for Particle-Based
  Materials Modeling at the Atomic, Meso, and Continuum Scales. \emph{Computer
  Physics Communications} \textbf{2022}, \emph{271}, 108171\relax
\mciteBstWouldAddEndPuncttrue
\mciteSetBstMidEndSepPunct{\mcitedefaultmidpunct}
{\mcitedefaultendpunct}{\mcitedefaultseppunct}\relax
\EndOfBibitem
\bibitem[Walker \latin{et~al.}(2014)Walker, Bryson, Hayward, and
  Tew]{walkerWide2014}
Walker,~C.~N.; Bryson,~K.~C.; Hayward,~R.~C.; Tew,~G.~N. Wide {{Bicontinuous
  Compositional Windows}} from {{Co-Networks Made}} with {{Telechelic
  Macromonomers}}. \emph{ACS Nano} \textbf{2014}, \emph{8}, 12376--12385\relax
\mciteBstWouldAddEndPuncttrue
\mciteSetBstMidEndSepPunct{\mcitedefaultmidpunct}
{\mcitedefaultendpunct}{\mcitedefaultseppunct}\relax
\EndOfBibitem
\bibitem[Auhl \latin{et~al.}(2003)Auhl, Everaers, Grest, Kremer, and
  Plimpton]{auhlEquilibration2003}
Auhl,~R.; Everaers,~R.; Grest,~G.~S.; Kremer,~K.; Plimpton,~S.~J. Equilibration
  of Long Chain Polymer Melts in Computer Simulations. \emph{The Journal of
  Chemical Physics} \textbf{2003}, \emph{119}, 12718--12728\relax
\mciteBstWouldAddEndPuncttrue
\mciteSetBstMidEndSepPunct{\mcitedefaultmidpunct}
{\mcitedefaultendpunct}{\mcitedefaultseppunct}\relax
\EndOfBibitem
\bibitem[Sliozberg and Andzelm(2012)Sliozberg, and Andzelm]{sliozbergFast2012}
Sliozberg,~Y.~R.; Andzelm,~J.~W. Fast Protocol for Equilibration of Entangled
  and Branched Polymer Chains. \emph{Chemical Physics Letters} \textbf{2012},
  \emph{523}, 139--143\relax
\mciteBstWouldAddEndPuncttrue
\mciteSetBstMidEndSepPunct{\mcitedefaultmidpunct}
{\mcitedefaultendpunct}{\mcitedefaultseppunct}\relax
\EndOfBibitem
\bibitem[Ge \latin{et~al.}(2013)Ge, Pierce, Perahia, Grest, and
  Robbins]{geMolecular2013}
Ge,~T.; Pierce,~F.; Perahia,~D.; Grest,~G.~S.; Robbins,~M.~O. Molecular
  Dynamics Simulations of Polymer Welding: {{Strength}} from Interfacial
  Entanglements. \emph{Physical Review Letters} \textbf{2013}, \emph{110},
  1--5\relax
\mciteBstWouldAddEndPuncttrue
\mciteSetBstMidEndSepPunct{\mcitedefaultmidpunct}
{\mcitedefaultendpunct}{\mcitedefaultseppunct}\relax
\EndOfBibitem
\bibitem[{Widmer-Cooper} \latin{et~al.}(2004){Widmer-Cooper}, Harrowell, and
  Fynewever]{widmer-cooperHow2004}
{Widmer-Cooper},~A.; Harrowell,~P.; Fynewever,~H. How {{Reproducible Are
  Dynamic Heterogeneities}} in a {{Supercooled Liquid}}? \emph{Physical Review
  Letters} \textbf{2004}, \emph{93}, 135701\relax
\mciteBstWouldAddEndPuncttrue
\mciteSetBstMidEndSepPunct{\mcitedefaultmidpunct}
{\mcitedefaultendpunct}{\mcitedefaultseppunct}\relax
\EndOfBibitem
\bibitem[Colombo \latin{et~al.}(2013)Colombo, {Widmer-Cooper}, and
  Del~Gado]{colomboMicroscopic2013}
Colombo,~J.; {Widmer-Cooper},~A.; Del~Gado,~E. Microscopic {{Picture}} of
  {{Cooperative Processes}} in {{Restructuring Gel Networks}}. \emph{Physical
  Review Letters} \textbf{2013}, \emph{110}, 198301\relax
\mciteBstWouldAddEndPuncttrue
\mciteSetBstMidEndSepPunct{\mcitedefaultmidpunct}
{\mcitedefaultendpunct}{\mcitedefaultseppunct}\relax
\EndOfBibitem
\bibitem[Hagberg \latin{et~al.}(2008)Hagberg, Schult, and
  Swart]{SciPyProceedings_11}
Hagberg,~A.~A.; Schult,~D.~A.; Swart,~P.~J. Exploring Network Structure,
  Dynamics, and Function Using {{NetworkX}}. Proceedings of the 7th Python in
  Science Conference. {Pasadena, CA USA}, 2008; pp 11--15\relax
\mciteBstWouldAddEndPuncttrue
\mciteSetBstMidEndSepPunct{\mcitedefaultmidpunct}
{\mcitedefaultendpunct}{\mcitedefaultseppunct}\relax
\EndOfBibitem
\bibitem[Brandes(2001)]{brandesFaster2001}
Brandes,~U. A Faster Algorithm for Betweenness Centrality. \emph{The Journal of
  Mathematical Sociology} \textbf{2001}, \emph{25}, 163--177\relax
\mciteBstWouldAddEndPuncttrue
\mciteSetBstMidEndSepPunct{\mcitedefaultmidpunct}
{\mcitedefaultendpunct}{\mcitedefaultseppunct}\relax
\EndOfBibitem
\bibitem[Brandes(2008)]{brandesVariants2008}
Brandes,~U. On Variants of Shortest-Path Betweenness Centrality and Their
  Generic Computation. \emph{Social Networks} \textbf{2008}, \emph{30},
  136--145\relax
\mciteBstWouldAddEndPuncttrue
\mciteSetBstMidEndSepPunct{\mcitedefaultmidpunct}
{\mcitedefaultendpunct}{\mcitedefaultseppunct}\relax
\EndOfBibitem
\bibitem[de~Gennes and Prost(1993)de~Gennes, and Prost]{gennesPhysics1993}
de~Gennes,~P.~G.; Prost,~J. \emph{The {{Physics}} of {{Liquid Crystals}}};
  {Clarendon Press}, 1993\relax
\mciteBstWouldAddEndPuncttrue
\mciteSetBstMidEndSepPunct{\mcitedefaultmidpunct}
{\mcitedefaultendpunct}{\mcitedefaultseppunct}\relax
\EndOfBibitem
\end{mcitethebibliography}
\providecommand{\latin}[1]{#1}
\makeatletter
\providecommand{\doi}
  {\begingroup\let\do\@makeother\dospecials
  \catcode`\{=1 \catcode`\}=2 \doi@aux}
\providecommand{\doi@aux}[1]{\endgroup\texttt{#1}}
\makeatother
\providecommand*\mcitethebibliography{\thebibliography}
\csname @ifundefined\endcsname{endmcitethebibliography}
  {\let\endmcitethebibliography\endthebibliography}{}

\end{document}